# Advancing 1D thermo-hydraulic tools for large cryogenic facilities

**R Beckwith, R Bruce, S Koshelev**

Fermi National Accelerator Laboratory, Batavia, Illinois 60510, USA

Email: rbruce@fnal.gov

**Abstract**. The Cryogenic Division at Fermilab develops large-scale cryogenic systems for particle accelerators and superconducting magnet test facilities. To support design and diagnostics, a Python-based code was created to calculate pressure drops in components such as valves and pipes. This paper presents recent enhancements to the code, including new heat transfer functions that improve the accuracy of thermal and hydraulic predictions. The first application models pressure and temperature changes in Proton Improvement Plant PIP-II relief pipes, aiding pipe sizing and protecting relief valves. The second example analyses heat load evolution in a pipe carrying sub-atmospheric helium, helping interpret temperature sensor data and understand gas return behavior to cold compressors. These improvements significantly expand the tool's capabilities, offering a practical resource for designing and troubleshooting cryogenic systems under dynamic thermal and flow conditions.

## 1. Introduction

Designing cryogenic systems for large-scale particle accelerators requires accurate modeling of pressure drops and heat loads across complex piping networks. At Fermilab, a Python-based computational library called CryoToolbox [1] was developed to support this effort by providing hydraulic analysis in cryogenic lines containing components such as elbows, tees, manual valves, control valves, and relief devices. Originally focused on pressure drop calculations, the tool has been significantly enhanced to include heat transfer functions, enabling more complete thermo-hydraulic evaluations of cryogenic systems. The library automatically handles unit conversions and fluids properties. It also supports flexible material property inputs, either by using a database of standard cryogenic materials or through manual specification.

    The primary goal of this paper is to present these recent developments, which expand the tool's capability from pressure drop analysis to include heat transfer under various conditions. The first part of this paper provides an overview of the underlying friction factor and pressure drop formulations, followed by the introduction of the new heat transfer models. These include forced convection correlations for heated pipe sections, as well as additional functions that estimate heat input through multilayer insulation (MLI) or frozen segments. The second section introduces the newly implemented thermal analysis features. Finally, two application examples illustrating both design and diagnostic use cases are presented. The first example presents the pressure and temperature evolution in the bayonet can relief piping of the Proton Improvement Plant PIP-II under various heating conditions. The second case shows the heat load analysis in long sub-atmospheric helium transfer lines. Together, these examples demonstrate the library's utility in both design and diagnostics of cryogenic systems.

## 2. The Cryotoolbox python library

The first section will present the existing toolbox and the heat transfer functions that have been added to the current version of the code.

*2.1. The CryoToolBox python library*

CryoToolbox is a Python-based library developed to support the modeling, analysis, and design of cryogenic systems, with particular emphasis on applications involving transfer lines, cryogenic heat exchangers, and low-temperature thermo-hydraulic phenomena. Tailored for scientific and engineering use, the library offers a comprehensive suite of computational tools and material property databases to perform thermophysical property evaluation, pressure drop estimation, heat transfer calculations on large cryogenic system. CryoToolbox integrates seamlessly with Python's scientific ecosystem and is designed with a modular architecture, allowing users to easily extend its capabilities. It serves as a practical and flexible resource for researchers and engineers working on cryogenic technologies in accelerator systems, space applications, or fundamental low-temperature research.

*2.1.1. Friction factor and pressure drop calculation*

The CryoToolbox library estimates pressure drops in cryogenic flow systems using a combination of empirical correlations and engineering models tailored to a wide range of Reynolds numbers and pipe surface roughness. For single-phase flow, calculations are based on the Darcy-Weisbach equation, with the friction factor determined according to the flow regime. Users can select from several established correlations, including the Modified Churchill, Serghide, and Zigrang-Sylvester methods [2], providing flexibility for different applications and flow conditions. Local pressure losses due to fittings—such as valves, elbows, and contractions—are incorporated through loss coefficients (K-values) derived from cryogenics-relevant literature and standard engineering references [3]. To ensure accurate thermophysical property inputs, fluid properties such as density and viscosity are dynamically retrieved from external databases including HEPAK, CoolProp, and RefProp. This approach enables robust and precise modeling of pressure behavior in low-temperature fluid systems.

*2.1.2. Components and line structure or the Line class*

In the CryoToolbox library, the Line class provides a structured framework for connecting multiple components—such as valves, elbows, and pipes of varying diameters—into a continuous cryogenic flow path (see Fig. 1). This modular approach enables the calculation of the total pressure drop across the entire system by sequentially evaluating the hydraulic and thermal effects of each individual segment. For each component, the computed outlet pressure and enthalpy are used as the inlet conditions for the next, ensuring accurate propagation of thermodynamic states along the line. When two adjacent components differ in diameter, the class automatically accounts for additional pressure losses due to sudden expansions or contractions, using appropriate loss coefficient models. This feature allows for more realistic modelling of complex piping systems with geometric discontinuities and varying flow regimes.

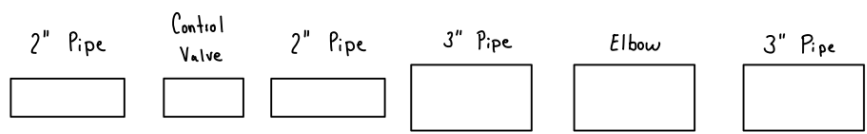

**Figure 1.** Schematic representing the Line class

Beyond system-level analysis, the Line class also supports segmented modeling to enhance the accuracy of pressure drop calculations. The Darcy-Weisbach equation, commonly used for this purpose, assumes incompressible flow—a valid approximation when the pressure drop is less than approximately 10% of the inlet pressure. However, for cryogenic applications involving larger pressure variations, compressibility effects become significant, and the incompressible assumption no longer holds. To address this limitation, the Line class automatically divides long transfer lines into smaller segments, ensuring that the pressure drop across each segment remains within the 10% threshold. This segmented strategy allows for the stepwise recalculation of pressure, temperature, and fluid properties along the flow path, thereby improving the accuracy of thermo-hydraulic predictions in compressible flow regimes (see Fig. 2).

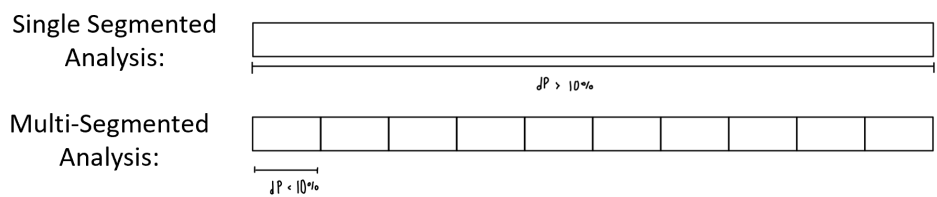

**Figure 2.** Schematic representing the Line class

*2.2. Heat Transfer Function Updates*
To more accurately evaluate pressure drops in large cryogenic facilities, the library accounts for the evolution of fluid temperature by incorporating the effects of heat loads applied to individual components. To support this, both internal and external heat transfer coefficients have been implemented in the code, enabling the simulation of heat exchange between the fluid and its surroundings throughout the flow path.

*2.2.1. Heat transfer functions*
The CryoToolbox library includes internal convective heat transfer calculations for flow within tubes and pipes. The internal heat transfer coefficient is computed using classical Nusselt number correlations from Nellis and Klein [4], applicable across a wide range of flow regimes and geometries. For laminar flow, the library employs empirical relations derived from Bennett, while for turbulent flow, it implements the Gnielinski correlation, which incorporates the friction factor. Users can select from several friction factor models described in Section 2.1 to best match their application. All heat transfer functions are supported by unit tests and have been cross validated against results from the Engineering Equation Solver (EES software) to ensure numerical consistency and reliability across regimes.
   External heat transfer is also addressed within the library. For natural convection around cylindrical geometries, the Churchill correlation is used [5]. In cases where the pipe is placed inside an insulating vacuum space and wrapped with multilayer insulation (MLI), the library implements the most recent correlations reviewed by Singh et al. [6]. This model accounts for the number of MLI layers, layer density, and vacuum pressure—allowing users to evaluate the thermal impact of degraded vacuum conditions on the performance of cryogenic components.

*2.2.2. The icing function*
During vacuum loss events in cryogenic systems or during the quench of superconducting magnets, large volumes of cold helium are rapidly released. In many cases, this helium is vented through non-insulated piping, in which its external surface is exposed to ambient air. As a result, these surfaces can freeze and become covered with ice. The formation of an ice layer significantly alters the external heat transfer characteristics by decreasing the heat exchange between the pipe and environment. To more accurately capture this effect, a dedicated heat transfer model for ice-covered surfaces has been implemented in the CryoToolbox library, based on the following correlation []:

$〚Lewis〛\_Number = Schmidt/Prandtl$ & $〚Lewis〛\_Relationship = 〚Lewis\ Number〛^{(-2/3)}/(\rho * C\_p)$
$〚Mass\ flux〛\_condensation = h\_convection * 〚Lewis〛\_Relationship * \Delta Concentration$

$〚Heat\ flux〛\_condensation = 〚Mass\ flux〛\_condensation * \Delta Enthalpy$

The CryoToolbox library calculates a modified heat transfer coefficient that accounts for the energy required to form ice on the pipe surface. This formulation incorporates the local dew point to estimate the onset of condensation and freezing. Currently, the model assumes a negligible ice thickness, as it is primarily intended for short-duration transient events, such as rapid helium discharge. However, future versions of the library may include ice thickness as a user-defined parameter to enhance accuracy for longer-duration scenarios.

## 3. Heat transfer Functions

Building on the heat transfer functions described in the previous section, this section of the paper presents the methods implemented in the CryoToolbox library to automatically calculate the temperature evolution of both the fluid inside the pipe and the pipe wall itself under various thermal conditions.

### 3.1. Flow evolution with a fixed heat load and temperature

The CryoToolbox library implements two methods to evaluate fluid temperature evolution in a pipe, depending on external thermal boundary conditions.

In the first method, a constant heat flux is applied to the outer surface of the pipe (see Fig. 3 – left). The code computes the average fluid temperature based on the enthalpy change between the inlet and outlet. Using this average temperature, it calculates the pressure drop and internal heat transfer coefficient, as described in the previous sections. To determine the inner wall temperature, the method employs an optimization routine from the SciPy library, which has been integrated into CryoToolbox. The outer wall temperature is then computed by integrating the thermal conductivity through the pipe wall, using the material's temperature-dependent properties.

The second method assumes a constant external wall temperature (see Fig. 3 – right). In this configuration, the average fluid temperature is computed iteratively. The first iteration assumes the inlet temperature to estimate the heat transfer and outlet temperature. The next iteration uses the newly calculated outlet temperature to refine the average temperature of the fluid. This loop continues until the difference between successive average temperatures of the fluid falls below 1 mK, ensuring convergence.

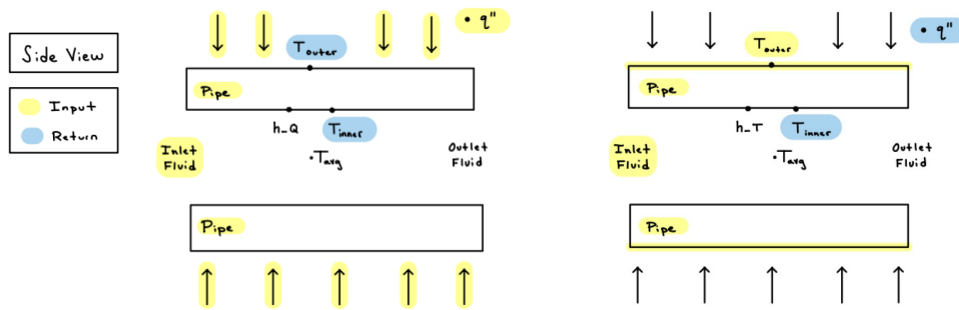

**Figure 3.** Schematic representing the method to calculate the temperature evolution with a constant heat load (left) and uniform temperature (right) on the external surface of the pipe

### 3.2. Flow evolution with a fixed outer heat transfer coefficient

This final method assumes a fixed heat transfer coefficient on the outer surface of the pipe. Similar to the approach with a constant heat flux, the inner wall temperature is determined by first calculating the external heat load using the specified outer heat transfer coefficient. The corresponding heat transfer is then used to solve for the inner wall temperature using the equation below:

$$T_{inner\ wall} = \left(\frac{h_{external}}{h_T}\right) * \left(\frac{D_o}{D_i}\right) * (T_{external} - T_{outer\ wall}) + T_{average}$$

An iterative method is employed to more accurately estimate the average fluid temperature within the pipe, which in turn improves the calculation of heat transfer between the fluid and the pipe wall.

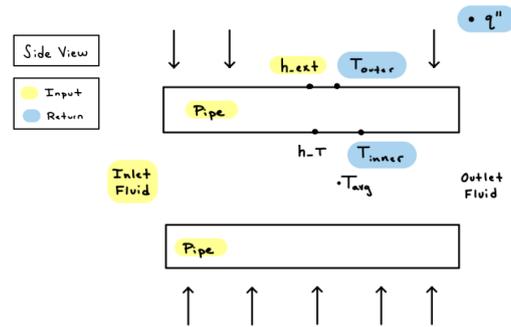

**Figure 4.** Schematic representing the method to calculate the temperature evolution with a fixed outer heat transfer coefficient on the external surface of the pipe

In this method, the fixed external heat transfer coefficient can be replaced by either natural convection correlations for cylindrical surfaces or the "icing" correlation described in the previous sections. This flexibility allows for a more accurate prediction of pressure and temperature evolution within the fluid under varying thermal boundary conditions.

## 4. Examples using the CryotoolBox library

This section presents two application examples demonstrating how the newly implemented heat transfer functions in the CryoToolbox library enhance the estimation of temperature and pressure evolution, as well as heat loads, in cryogenic transfer lines.

### 4.1. PIP-II Bayonet Can Relief Piping

The first application of the new functions in the CryoToolbox library is to estimate the temperature and pressure evolution within the relief piping system that protects the bayonet cans along the tunnel transfer line of PIP-II.

#### 4.1.1. The PIP-II relief system

The relief piping that protects the bayonets of the tunnel transfer line consists of a single pipe, with one section thermally insulated within a vacuum space and another section exposed to ambient conditions at atmospheric pressure and temperature (see Fig. 5). The objective of this analysis is to estimate the potential overpressure that may develop at the onset of a relief event—specifically, when the relief valve just begins to open. Additionally, the calculation aims to verify whether valve chatter may occur during the discharge. To prevent chatter, the upstream pressure must remain below 3% of the set relief pressure once the valve begins to open.

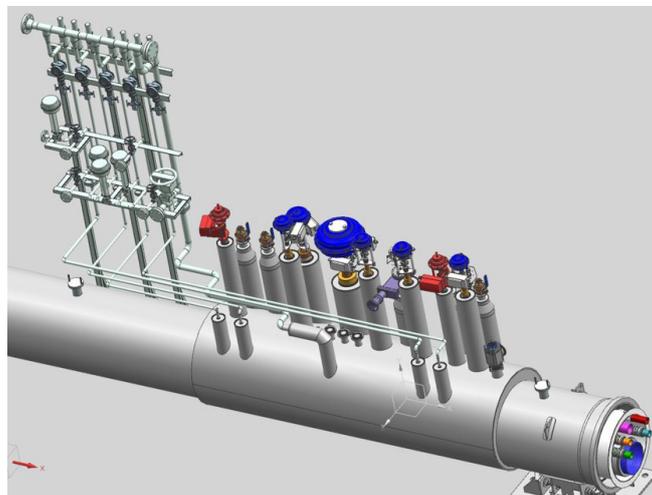

**Figure 5.** Design of PIP II tunnel transfer line with the relief piping

*4.1.2. Pressure and temperature evolution in these lines.*
This study presents results for the DN40 line that supplies helium to the thermal shield of the tunnel transfer line. During a vacuum break event, the shield may experience significant heat loads, leading to a maximum estimated helium mass flow of 258 g/s. Based on the 3% rule for relief systems, the allowable pressure drop upstream of the relief valve is limited to 720 mbar, assuming a valve set pressure of 24 bar. Pressure drop calculations were performed under various heating scenarios.

In the first scenario, the pipe's external surface was held at a fixed temperature, representing the initial phase of the relief event when the non-insulated section is still at ambient temperature. This approach assesses whether the initial overpressure remains within the allowable design limits of the piping. Three outer wall temperatures were analyzed (see Fig. 6). In the worst-case scenario—an outer wall temperature of 293 K on the non-insulated section—the total pressure drop was 2.84 bar, keeping the system pressure below the allowable limit of 30 bar.

A second set of simulations considered convective heat transfer, using external heat transfer coefficients up to 10 W/(m²·K) (typical ambient estimates are around 2 W/(m²·K)) and uniform heat fluxes up to 10 kW/m². In all cases, the upstream pressure remained below the 720 mbar threshold, indicating no risk of relief valve chatter.

| Inlet Parameters: | | |
|---|---|---|
| mass flow rate | g/s | 258 |
| P inlet (for internal piping) | bar | 29.04 |
| P inlet (for external piping) | bar | 28.15 |
| T inlet | K | 40 |
| 3% Pressure Drop | mbar | 720 |

| | | No Applied Heat | Defined Outer Wall Temperature | | | Defined External Heat Transfer Coefficient | | | Defined Applied Heat Flux | | |
|---|---|---|---|---|---|---|---|---|---|---|---|
| | | | 293K | 250K | 200K | 1 W/(K*m^2) | 5 W/(K*m^2) | 10 W/(K*m^2) | 500 W/m^2 | 1000 W/m^2 | 10000 W/m^2 |
| Parameter | Unit | | | | | | | | | | |
| dP (external piping) | mbar | 610.4 | 2840.2 | 2350.9 | 1812.2 | 611.2 | 614.5 | 618.5 | 611.8 | 613.1 | 637.7 |
| Outlet Temp | K | 40 | 249.7 | 209.4 | 163.5 | 40.1 | 40.6 | 41.1 | 40.2 | 40.4 | 43.8 |
| Inner Wall Temp Range | K | 40-40 | 103-263.8 | 85.2-222 | 65.6-174 | 40-40.1 | 40.1-40.6 | 40.2-40.3 | 40-40.2 | 40.1-40.5 | 41-44.6 |
| Outer Wall Temp Range | K | 40-40 | 293-293 | 250-250 | 200-200 | 40.1-40.2 | 40.7-41.3 | 41.4-42.5 | 40.3-40.5 | 40.6-40.9 | 45.5-48.8 |
| Heat Flux Range on Each Component | W/m^2 | 0-0 | 1110000-223000 | 987000-197000 | 721000-165000 | 300.3-300.2 | 1498-1495 | 2988-2975 | 500-500 | 1000-1000 | 10000-10000 |

**Figure 6.** Summary of Results for one relief line DN40

*4.2. Sub-atmospheric transfer Line heat loads*
In a second application, the CryoToolbox library was used to analyse the evolution of heat loads in sub-atmospheric helium transfer lines.

*4.2.1. The heat load estimation issue*
Several particle accelerators, such as PIP-II and LCLS-II, rely on saturated superfluid helium baths operating below 2 K and at pressures under 30 mbar. These baths are connected to their respective compressors through long sub-atmospheric transfer lines, and accurately characterizing the evolution of heat loads along these lines is essential for the stable operation of these large-scale cryogenic systems. In practice, sensors used to estimate heat loads are typically mounted on the outer surface of the piping. However, under sub-atmospheric conditions, the internal heat transfer coefficient is extremely low, which can lead to a significant temperature difference between the fluid inside the pipe and the sensor's reading. To address this challenge, the CryoToolbox library implements an iterative method that estimates a more realistic heat load by reconciling the known bath temperature (determined from the saturated pressure) with the temperature measured at the sensor location (see Fig. 7). This approach improves the accuracy of heat load estimations in sub-atmospheric helium systems and enhances diagnostic and operational capabilities.

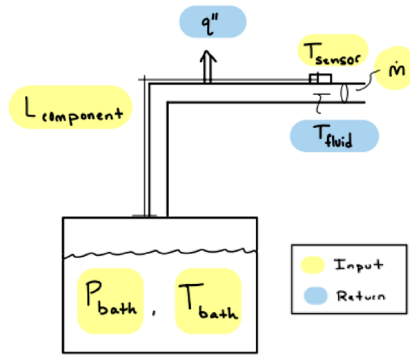

**Figure 7.** Schematic of the sub-atmospheric transfer line inputs and outputs

*4.2.2. The estimated heat load error*
A test case was conducted using a 250 mm diameter pipe (10-inch Schedule 10), through which sub-atmospheric helium gas flows at 2 K and 30 mbar with a mass flow rate of 100 g/s. These parameters are representative of the conditions in the LCLS-II sub-atmospheric transfer line currently operating at SLAC [8]. To better capture the temperature evolution of the fluid, the pipe was divided into 20 segments. A uniform heat load was assumed on the outer surface of the pipe, and a temperature sensor was placed at a distance of 8.25 meters from the helium bath. For this analysis, a sensor temperature of 2.1 K was arbitrarily chosen to estimate the heat load.

Traditionally, the measured surface temperature is used directly to estimate the heat load at that section. Applying this conventional method to the example yields a heat load of 8.29 W/m². In contrast, the iterative approach implemented in the CryoToolbox library begins by using this initial estimate to calculate the expected temperature difference between the fluid and the pipe wall at the sensor location. This difference is then subtracted from the sensor reading to approximate the actual fluid temperature, which is used to recalculate the heat load. After convergence, the updated heat load is 1.25 W/m²—more than six times lower than the original estimate.

Although the parameters in this example are not drawn from a specific operational system, the results highlight how the new method can significantly improve the accuracy of heat load estimation in cryogenic transfer lines, particularly in low-pressure conditions where the internal heat transfer coefficient is very low.

## 5. Conclusion
Recent enhancements to the CryoToolbox library have significantly improved its capabilities for modeling pressure drop and heat transfer in cryogenic piping systems. By integrating validated correlations for internal convection, external natural convection, multilayer insulation (MLI), and surface icing, the library offers more accurate tools for predicting temperature and pressure evolution under steady-state conditions. The addition of iterative and optimization-based methods further increases accuracy in scenarios involving complex thermal boundary conditions, such as fixed wall temperature or uniform heat flux.

Two application examples highlight the extended capabilities of the library. The first demonstrates how the new features can be used to assess overpressure risks and verify valve performance in the relief piping of the PIP-II transfer line. The second example illustrates how conventional surface temperature measurements can lead to significant overestimation of heat loads in sub-atmospheric helium lines, and how the iterative approach implemented in CryoToolbox provides more reliable estimates under low-convection conditions.

Future updates will continue to expand its functionality, including the addition of transient modelling capabilities, to support the analysis and design of increasingly complex cryogenic systems.

**Acknowledgments**

This manuscript has been authored by FermiForward Discovery Group, LLC under Contract No. 89243024CSC000002 with the U.S. Department of Energy, Office of Science, Office of High Energy Physics. FERMILAB-CONF-25-0329-TD